\documentclass[nofootinbib,aps,prl,a4paper,twocolumn,english,superscriptaddress,longbibliography,reprint]{revtex4-2}
\usepackage[mathlines]{lineno}

\usepackage{amsthm}
\usepackage{amsmath}
\usepackage{amssymb}
\usepackage{graphicx}
\usepackage{bm}
\usepackage{color}
\usepackage[dvipsnames]{xcolor}
\usepackage{enumitem}
\usepackage{mathrsfs}
\usepackage{verbatim}
\usepackage{bbold}
\usepackage{braket}
\usepackage{lipsum}
\usepackage{url}
\usepackage{tikz}
\usepackage{graphicx}
\usepackage[colorlinks=true,citecolor=blue,urlcolor=blue]{hyperref}

\newcommand{\Tr}{\text{Tr}}
\newcommand{\ave}[1]{\langle #1 \rangle}
\newcommand{\cl}[1]{\hat{\mathcal{#1}}}

\usepackage[normalem]{ulem}

\begin{document}   
\title{Non-equilibrium dissipative stabilization of  s- and d-wave superconductivity}
\author{Aleksey Lunkin}
\affiliation{Nanocenter CENN, 1000 Ljubljana, Slovenia}
\author{Yury Holubeu}
\affiliation{Nanocenter CENN, 1000 Ljubljana, Slovenia}
\author{Denis Gole\v{z}}
\affiliation{Jo\v{z}ef Stefan Institute, 1000 Ljubljana, Slovenia}
\affiliation{Faculty of Mathematics and Physics, University of Ljubljana, 1000 Ljubljana, Slovenia}
\author{Zala Lenar\v{c}i\v{c}}
\affiliation{Jo\v{z}ef Stefan Institute, 1000 Ljubljana, Slovenia}

\begin{abstract} 
Stabilizing superconductivity beyond its equilibrium regime has been one of the quests of nonequilibrium state engineering. We report the presence of nonthermal pairing gap in BCS superconductors coupled to two thermal baths, where one of the baths can be at a temperature well above the equilibrium $T_c$. Superconducting state is enabled by a highly non-equilibrium steady state stabilized by weak coupling to baths that break detailed balance condition. The steady-state is well described by a generalized Gibbs ensemble parametrized with chemical potentials for associated Bogoliubov quasiparticle. A strong, non-perturbative effect is reported for both $s$-wave and $d$-wave superconductors, and its possible realization in cavity-based experiments is discussed.
\end{abstract}

\maketitle

Symmetry-broken states exhibit fascinating quantum many-body effects—including superconductivity, charge order, and quantum magnetism—realized in solid-state systems~\cite{keimer2017} and quantum simulators \cite{greiner02,bloch12,mazurenko17}. Traditionally, these phases were considered strictly dictated by system's intrinsic equilibrium properties. However, remarkable experimental progress enables researchers to engineer ordered phases via novel platforms~\cite{landig16,zhang17,mivehvar2021,Young2024,Young25,chu2025} or use external stimuli to stabilize non-thermal states~\cite{Giannetti2016,Basov2017,Sentef2021,murakami2025}. Examples include light-induced insulator-metal transitions~\cite{Perfetti2006,Perfetti2006,Okamoto07}, hidden phases in $1\text{T-TaS}_2$~\cite{Stojchevska2014,Vaskivskyi2016}, tellurides~\cite{Kogar2019}, and light-induced superconductivity~\cite{Mitrano2019,Fausti11,yursa2026phase}. A major limitation of these active driving approaches is that external sources inject large amounts of energy~\cite{murakami2017,Giannetti2016}. The resulting strong heating restricts applicability across broader physical systems, necessitating alternative manipulation strategies.

An emergent strategy for control is cavity engineering~\cite{ritsch2013cold,basov2025polaritonic,schlawin2022cavity}, where modifying electromagnetic modes enables selective coupling to active degrees of freedom, remarkably altering individual quantum levels and collective states of matter~\cite{purcell1946,Haroche01,haroche2006,kimble2008,ritsch2013cold,lukin2003}. For instance, recent experiments demonstrate cavity-altered insulator-metal transitions in $1\text{T-TaS}_2$~\cite{Jarc2023}, alongside modifications of superconductivity~\cite{Keren2026,xu2026,montanaro2026,zhang2026}, transport~\cite{kumar2024}, and topological properties~\cite{Appugliese2022}. While detailed descriptions remain debated, the standard theoretical framework relies on modified electromagnetic fluctuations~\cite{schlawin2022cavity,flick2017atoms}. These fluctuations induce new pairing channels predicted to alter ordered phases like superconductivity~\cite{sentef2018cavity,li2020,gao2020,curtis2019,schlawin2019,zachary2020} or ferroelectricity~\cite{lenk2022,ashida2020,latini2021,mazza2019,lenk2020}. However, this approach is currently constrained by weak coupling regimes and long wavelengths limitations. Overcoming these bottlenecks is a focus of intense research, using smart designs to achieve stronger light-matter coupling~\cite{basov2025polaritonic,rabl2018,schlawin2019,forn2019,GarciaVidal2021,Lu2025}.

However, engineering the electromagnetic environment induces also nonequilibrium conditions, opening the avenue of dissipative engineering~\cite{islam2025,curtis2019,Jarc2023,flores2025,fassioli2025}. For superconductors, previous analyses have primarily relied on field-theoretic approaches, such as generalized Eliashberg theory~\cite{eliashberg1970,ivlev1970influence,ivlev1973}, in which coupling to an electromagnetic bath induces weak deviations from equilibrium~\cite{curtis2019}. These non-thermal populations can, in turn, modify the pairing interaction~\cite{pini2026} or lift the degeneracy between superconductivity and charge-density-wave order~\cite{islam2025}. Similar effects occur in Floquet-driven BCS superconductors~\cite{Huanyu25}, though these suffer from strong heating.

\begin{figure}[t!]    
\includegraphics{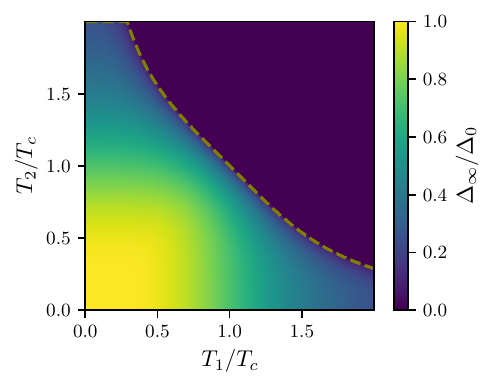}
\caption{Steady-state gap $
\Delta_\infty$ (in units of zero temperature gap $\Delta_0$) as a function of temperatures of the first bath $T_1$ and second bath $T_2$. For an s-wave superconductor, we find new non-equilibrium solutions at $T_1\neq T_2$ with finite gap despite one of the baths being at a temperature larger than the critical temperature $T_c$ in the equilibrium. The olive line represents the second-order phase-transition line.}\label{figure:fig1}
\end{figure}

An appealing, largely unexploited property of steady states manipulation is the ``greenhouse effect'' \cite{lange17}: at a weak system-environment coupling, constant driving and dissipation can yield an $O(1)$ effect if the system possesses nearly conserved quantities that become non-thermally (strongly) populated by driving~\cite{lange17,lange18,ulcakar23,ulcakar24}. In systems with quasi-particle occupations as (approximate) conserved quantities, this implies reshuffling of quasiparticle content into possibly highly non-thermal distributions~
\cite{ulcakar24}. Previously demonstrated with nearly integrable systems where the notion of quasiparticles is inherited from integrability~\cite{lange17,lange18,reiter21,schmitt22,ulcakar23,ulcakar24,ulcakar2025,bouchoule20,rossini21,mazza22,gerbino23,perfetto23,rowlands23,riggio24,starchl22,starchl24, lumia24,lehr25,marche26}, similar non-thermal state engineering at weak driving could be realized in other models that admit a quasiparticle description, such as self-consistent mean-field treatments of interactions.

In this Letter, we demonstrate that the greenhouse effect applies to the dissipative engineering of both conventional ($s$-wave) and unconventional ($d$-wave) superconductors, leading to an $O(1)$ increase in the order parameter and extension of the superconducting phase despite weak system-bath coupling. We model the superconductor as a weakly dissipative system using an open-system extension of the BCS theory, capturing its non-equilibrium evolution via a time-dependent generalized Gibbs ensemble (tGGE) \cite{lange18,ulcakar24,ulcakar2025,bouchoule20,rossini21,mazza22,gerbino23,perfetto23,riggio24,lumia24,lehr25,marche26}. We consider a setup that corresponds to placing the material between two baths at temperatures $T_1$ and $T_2$. The resulting steady states significantly elevate $T_c$ for both pairing symmetries. Our main result is presented in Fig.~\ref{figure:fig1}, where the non-thermal $s$-wave order parameter is plotted against both temperatures, showing a finite value above $T_c$ if one bath is below it. 
Furthermore, we consider the effect of intrinsic scattering and possible experimental implementation of our proposal for dissipative stabilization of non-equilibrium superconductivity.

{\it Setup.} We consider a BCS superconductor coupled to two thermal baths at temperatures $T_1$ and $T_2$, described within the Lindblad master equation,
\begin{equation}\label{eq:liouv}
    \partial_t \rho = i[H(t), \rho] + \gamma_0 \mathcal{D}[\rho].
\end{equation}
The superconducting material is modelled by a BCS Hamiltonian
\begin{align}\label{eq:H0}
    H_0
 = \sum_{p}{}^{\prime} \, \big(&\xi_p (c^\dagger_{p,\uparrow}c_{p,\uparrow} + c^\dagger_{p,\downarrow}c_{p,\downarrow}) \\
 &- \sum_{p,p'}{}^{\prime} \,  \lambda(p,p') c^\dagger_{-p',\downarrow}c^\dagger_{p',\downarrow}c_{p,\uparrow} c_{-p,\uparrow},  \notag
\end{align}
on the mean field level simplified to a quadratic self-consistent problem
\begin{align}\label{eq:H}
    H(t)
 = \sum_{p}{}^{\prime} \,\big(&\xi_p (c^\dagger_{p,\uparrow}c_{p,\uparrow} + c^\dagger_{p,\downarrow}c_{p,\downarrow}) \\
 &+ \Delta_p(t) c^\dagger_{-p,\downarrow}c^\dagger_{p,\uparrow} + \Delta_p^*(t) c_{p,\uparrow} c_{-p,\downarrow} \big). \notag
\end{align}
Summation $\sum{}^{\prime}$ is limited by the cut-off on energies, $|\xi_p|<\omega_D$. In the following, we will use the zero temperature gap $\Delta_0$ as a natural energy unit which absorbs the cut-off and the interaction strength.

We consider weak coupling to baths $\gamma_0 
\ll \Delta_0$, implying slow changes of the state $\rho(t)$ on timescale $\gamma_0^{-1}$. Therefore, we assume that gap instantaneously follows the state
\begin{equation}
    \Delta_p(t) = -  \sum_{p'}{}^{\prime} \, \lambda(p,p') \Tr[c_{-p',\downarrow} c_{p',\uparrow} \rho(t)].
\end{equation}
Due to the self-consistency used in \eqref{eq:H}, Hamiltonian $H(t)$ is parametrized with a time-dependent gap function $\Delta_p(t)$, implying time-dependent quasiparticle creation and annihilation operators,
\begin{align}
\label{eq:quasiparticles}
    b_{p,\uparrow}(t) &= u_{p}(t) c_{p,\uparrow} + v_{p}(t) c_{-p,\downarrow}^\dagger, \\
    b_{p,\downarrow}(t) &= u_{p}(t)c_{p,\downarrow} - v_{p}(t)c_{-p,\uparrow}^\dagger, \nonumber\\
    u_{p}^2(t) &= \frac{1}{2}\left(1 + \frac{\xi_{p}}{\epsilon_{p}(t)}\right), \,  v_{p}^2(t) = \frac{1}{2}\left(1 - \frac{\xi_{p}}{\epsilon_{p}(t)}\right), \nonumber\\ 
    \epsilon_{p}(t) &=  \sqrt{\xi_p^2 + \Delta_p(t)^2}.\notag
\end{align}
In the following, we drop the time arguments. The action of thermal baths is represented by a sum of Lindblad terms that obey detailed balance at a given temperature,
$\cl{D}[\cdot] = \cl{D}^{(T_1)}[\cdot] + \cl{D}^{(T_2)}[\cdot]$,
\begin{align}
\label{eq:thermalD}
\mathcal{D}^{(T)}[\rho] &= 
\sum_{\substack{k_1, k_2 \\ s_1,s_2}}{}^{\hspace{-0.13cm}\prime} \, 
L^{(s_1,s_2)}_{k_1,k_2}\rho L^{(s_1,s_2)\dagger}_{k_1,k_2} - \frac{1}{2}\left\{\rho, L^{(s_1,s_2)\dagger}_{k_1 k_2}L^{(s_1,s_2)}_{k_1,k_2}\right\} \nonumber\\
L^{(s_1,s_2)}_{k_1,k_2} &= \sqrt{\gamma^{(s_1,s_2)}_{k_1,k_2}(T)} \, b^{(s_1)}_{k_1}b^{(s_2)}_{k_2}, 
\end{align}
with $b^{(+)}_{k} := b^{\dagger}_{k},  b^{(-)}_{k} := b_{k}$. We consider processes that cause creation of two quasiparticles $(s_1,s_2)=(+,+)$, annihilation of two quasiparticles $(s_1,s_2)=(-,-)$, and scattering between two particles $(s_1,s_2)=(-,+), (+,-)$ in order to avoid issues related to changes of parity sectors \cite{riggio24}. In our minimalistic modelling of the baths we only require them to obey the detailed balance at the given temperature, which posses conditions on rates $\gamma^{(s_1,s_2)}_{k_1,k_2}(T)$ that we discuss in the End Matter, where we also give a concrete form of the rates used.

{\it Method.}
Following previous works on weakly dissipative systems with well defined quasiparticles in the absence of dissipation \cite{ulcakar24,gerbino23,bouchoule20,riggio24}, we approximate the time-dependent density matrix as a generalized Gibbs ensemble with time dependent chemical potentials $\mu_{p,\sigma}(t)$ associated with quasiparticles of momentum $p$ and type $\sigma$,
\begin{equation}
    \rho(t) = \frac{e^{-\sum_{p,\sigma}\mu_{p,\sigma}(t) b_{p,\sigma}^\dagger(t)b_{p,\sigma}(t)}}{\Tr[e^{-\sum_{p,\sigma}\mu_{p,\sigma}(t) b_{p,\sigma}^\dagger(t)b_{p,\sigma}(t)}]}. 
\end{equation}
The intuition behind the GGE Anzatz is based on quasiparticles being a well defined concept on intermediate timescale, while the quasiparticle content is slowly reshuffled by the dissipators on long timescale $\gamma_0^{-1}$. At the same time we assume that the intermediate time scale is much longer than the short-time coherent dynamics~\cite{volkov73,yuzbashyan2006,Yuzbashyan2006b,Barankov2004,tsuji2015,kemper2015,blommel2025}. The redistribution of quasiparticle occupation is encoded in the time-dependence of chemical potentials $\mu_{p,\sigma}(t)$. The appropriateness of such approximation has been well tested (without self-consistency) for integrable systems that are weakly coupled to baths ~\cite{lange17,lange18,reiter21,schmitt22,ulcakar23,ulcakar24,bouchoule20,rossini21,mazza22,gerbino23,perfetto23,rowlands23,riggio24,starchl22,starchl24, lumia24,lehr25,marche26}. 

Due to the self-consistency, the equation of motion for the occupation numbers has two contributions
\begin{align}
    \partial_{t}n_{p,\sigma}(t) = &\Tr\Big[\partial_t \big(b_{p,\sigma}^\dagger (t)b_{p,\sigma}(t)\big) \, \rho(t)] \\
    &+ \Tr[ b_{p,\sigma}^\dagger (t) b_{p,\sigma}(t) \, \partial_t \rho(t)\Big]. \notag
\end{align}
However, using Eq.~(\ref{eq:quasiparticles}) one can show that the first term is exactly zero. The evolution equation thus reduces to
\begin{align}\label{eq:scattering}
      \partial_t n_{k}(t) = 2& \sum_{k^\prime}  \left[\Gamma^{(+,+)}_{k,k^\prime} (1 - n_{k}) (1 - n_{k^\prime}) -  \Gamma^{(-,-)}_{k,k^\prime} n_{k} n_{k^\prime}\right] \nonumber\\
     + &\sum_{k^\prime}\left[\Gamma^{(+,-)}_{k,k^\prime}(1 - n_{k})n_{k^\prime}     - \Gamma^{(+,-)}_{k^\prime,k}(1 - n_{k^\prime})n_{k} \right]  , \notag \\
     -& \gamma_1 [n_k - n_k^{\text{th}}],\\
     \Gamma^{(s_1,s_2)}_{k,k^\prime} = \phantom{1}&  \gamma^{(s_1,s_2)}_{k,k^\prime}(T_1) +  \gamma^{(s_1,s_2)}_{k,k^\prime}(T_2), 
\end{align}
where $k=\{\sigma, p\}$ and $\Gamma^{(s_1,s_2)}_{k,k^\prime}$ take into account two quasiparticle creation $(++)$, two quasiparticle annihilation $(--)$ and scattering $(+-/-+)$ generated due to coupling to the two baths. We assume equal coupling to the two baths for simplicity. 
In analogy with previous literature, we assume that the intrinsic scattering processes are treated within the relaxation time approximation at rate $\gamma_1$~\cite{ivlev1973,curtis2019}. 
Phenomenologically, the intrinsic interaction are pushing the distribution towards a thermal $n_k^{\text{th}}$ at the temperature that is in a dissipative setup given by the energy exchange with the environment \cite{lenarcic18,lenarcic20}.
We first neglect the intrinsic scattering ($\gamma_1=0$), since it is expected to be much weaker than any other energy scale $\gamma_1\ll\gamma_0\ll \Delta_0$~\cite{ivlev1973,curtis2019}. Later, we examine the opposite limit of strong thermalizing scattering processes. Interpolation between the two limits at finite intrinsic scattering is discussed in the Supplementary Material \cite{sm}.

The average quasiparticle occupation number within the GGE approximation is
$
n_{p,\sigma}(t) = 1/(1+ e^{\mu_{p,\sigma}(t)})
$
and the self-consistency condition for the gap becomes
\begin{equation}\label{eq:Delta_t}
    \Delta_p(t) =  \sum_{p'} \frac{\lambda(p,p')}{2 \epsilon_{p'}(t)} \left[1 - n_{p',\uparrow}(t)  - n_{p',\downarrow}(t) \right].
\end{equation} 
Our main goal is finding the steady state value of the gap function $\Delta_{p,\infty} = \lim_{t\to\infty} \Delta_p(t)$ that establishes at long times as a consequence of coupling to the two thermal baths. It can be obtained from a time propagation in which we initialize the quasiparticle distribution in some state, update the quasiparticle occupation numbers $n_{p,\sigma}(t)$ following Eq.~\eqref{eq:scattering} and adjust the gap $\Delta_p(t)$ at each time step according to Eq.~\eqref{eq:Delta_t} until it converges. Alternatively, steady-state occupations can be directly obtained by solving the steady-state self-consistency equation numerically or in some cases even analytically, see the End Matter for details on modelling. 

If the intrinsic scattering becomes so strong that the Bogolioubov modes can no longer be considered as partially conserved, evolution of quasiparticle occupations~\eqref{eq:scattering} is replaced with a single rate equation for the energy as the only approximately conserved quantity \cite{lenarcic18,lenarcic20}. In this case, the state is described with a single parameter - the time dependent temperature - calculated from the rate equation for the energy exchange with the baths, 
\begin{align}\label{eq:Hrate}
    \ave{\dot{H}}=\Tr\left[ H(t) \gamma_0 \cl{D} \frac{e^{-\beta(t) H(t)}}{\Tr[e^{-\beta(t) H(t)}]}\right],
\end{align}
which vanishes in the steady state.

Before proceeding to the results, it is worth comparing our formalism with previous Keldysh-based treatments for driven superconductors: (a) Our description based on GGE employs a Markovian description of thermal baths, imprinted via detailed-balance obeying weights in Eq.~\eqref{eq:gammas}, in contrast to the non-Markovian history integrals in Eliashberg treatment~\cite{curtis2019,ivlev1973,Eliashberg70,islam2025}. (b) 
The Markovian structure makes these equations considerably easier to solve, even analytically. Consequently, unlike in the conventional treatment of the Eliashberg effect, there is no need to linearize the equations~\cite{curtis2019,ivlev1973}, allowing one to describe large deviations from the thermal occupation.

\begin{figure}[t!]    
\includegraphics[width=0.9\columnwidth]{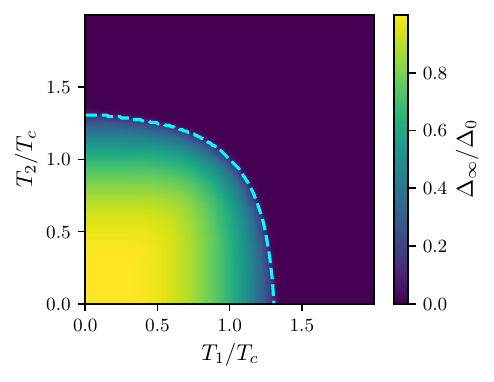}
\caption{
Steady-state gap as a function of temperatures of the first bath $T_1$ and second bath $T_2$ for an s-wave superconductor with strong inherent scattering $\gamma_1 \approx \gamma_0$, leading to a steady state well described by a temperature determined by the stationarity of energy, Eq.~\eqref{eq:Hrate}.}
\label{figure:fig1a}
\end{figure}

{\it Results for s-wave non-equilibrium superconductivity.}
We consider the behaviour at s-wave channel of the interaction, which reduces $\lambda(p,p')=\lambda$ and $\Delta_p=\Delta$.
In Fig.~\ref{figure:fig1}, we show the ratio between the steady state and zero-temperature gap  $\Delta_\infty/\Delta_0$ as a function of bath temperatures $T_1$ and $T_2$.   
We find that a non-equilibrium state at $T_1\neq T_2$ can support a finite gap in broad regions with $T_2 < T_c < T_1$ (and vice versa).
On the contrary, if the inherent scattering is strong $\gamma_1\approx \gamma_0$ and the steady state is thermal with a temperature determined from the energy rate equation \eqref{eq:Hrate}, the region with finite gap shrinks again, as shown in Fig.~\ref{figure:fig1a}

\begin{figure}[t!]    
\includegraphics[width=0.8\columnwidth]{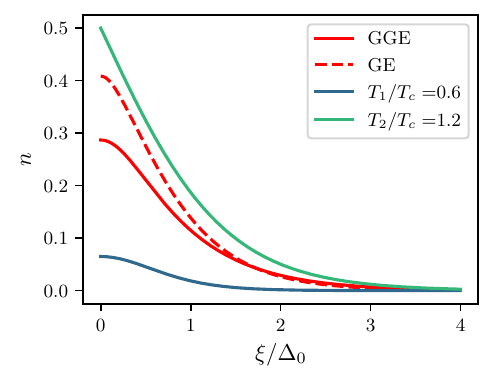}
\caption{
Different steady state distribution of quasiparticles at the bath temperatures $T_1/T_c = 0.6$ and $T_2/T_c = 1.2$. Red curve (GGE) corresponds to the non-equilibrium steady state, stabilized by the interplay of the two baths, Eq~\eqref{eq:scattering}. Dashed line (GE) represents the thermal occupation obtained from the energy rate equation \eqref{eq:Hrate} in case of strong inherent scattering $\gamma_1 \sim \gamma_0$. The blue and green curves correspond to thermal distributions at the temperatures of the two baths, reached if system would be coupled to only one of them.
}\label{figure:fig2}
\end{figure}

The new superconducting regions reported in Fig.~\ref{figure:fig1} thus stem from the non-thermal occupation of Bogoliubov modes, induced by weak coupling to the baths. Although each bath individually satisfies detailed balance and stabilizes a thermal state at its own temperature, the interplay of two baths at different temperatures can stabilize a non-trivial quasiparticle population that sustains superconductivity. In Fig.~\ref{figure:fig2}, we plot the steady-state occupation as a function of normalized  energy $\xi/\Delta_0$ for $T_1=0.6 T_c$ and $T_2=1.2 T_c$ (GGE). For comparison, we also show thermal distributions at the bath temperatures $T_1$ and $T_2$, alongside the thermal steady-state distribution (GE), obtained from the energy exchange equation \eqref{eq:Hrate} of relevance at a strong intrinsic scattering $\gamma_0\approx\gamma_1$.

Comparing these distributions connects our nonequilibrium superconductivity to the long-known Eliashberg effect~\cite{Eliashberg70,ivlev73} with quasiparticles being shifted to higher energies compared to the thermal distribution. 
However, our mechanism differs crucially in its parametric strength. The original Eliashberg mechanism relies on weak driving, and likewise, in its cavity extension proposed in Ref.~\cite{curtis2019}, the nonequilibrium response scales with the inelastic relaxation rate $\gamma_1$. In contrast, our mechanism produces an $\mathcal{O}(1)$ effect in the steady state. While the weak system–bath coupling $\gamma_0 \ll \Delta_0$ sets the slow relaxation timescale, it does not enter the steady-state condition $\partial_t n_k=0$ for the quasiparticle occupations in Eq.~\eqref{eq:scattering}.
Consequently, the manipulation of superconductivity remains exceptionally large even at weak coupling, as the deviation of the steady-state occupation from its thermal approximation is independent of $\gamma_0$ and determined entirely by the bath temperatures. Mechanism proposed in Ref.~\cite{pini2026} affects the pairing interaction which is complementary to our mechanism and it is an interesting future work to understand the interplay between the two.
In our setup, we observe a second order character of the transition along the whole non-equilibrium phase diagram. Expansion of the self-consistent steady state condition (End Matter) around the transition allows for analytical solution, confirming a $\Delta \sim \sqrt{\tilde{T}_{c} - T_1}$ dependence at a fixed $T_2$ and nonequilibrium critical $\tilde{T}\textbf{}_c$. 
This is in contrast to other Refs.~\cite{curtis2019,Huanyu25,pini2026}, achieving stabilization of non-equilibrium states by driving and dissipation, with an onset of superconductivity in a first order transition.
We speculate that a different nature of transition might be a consequence of Markovian baths considered, since all other Refs.~\cite{curtis2019,Huanyu25,pini2026} considered non-Markovian coupling. However, additional work is needed to confirm that.

\begin{figure}[t]    
\includegraphics[width=0.9\columnwidth]{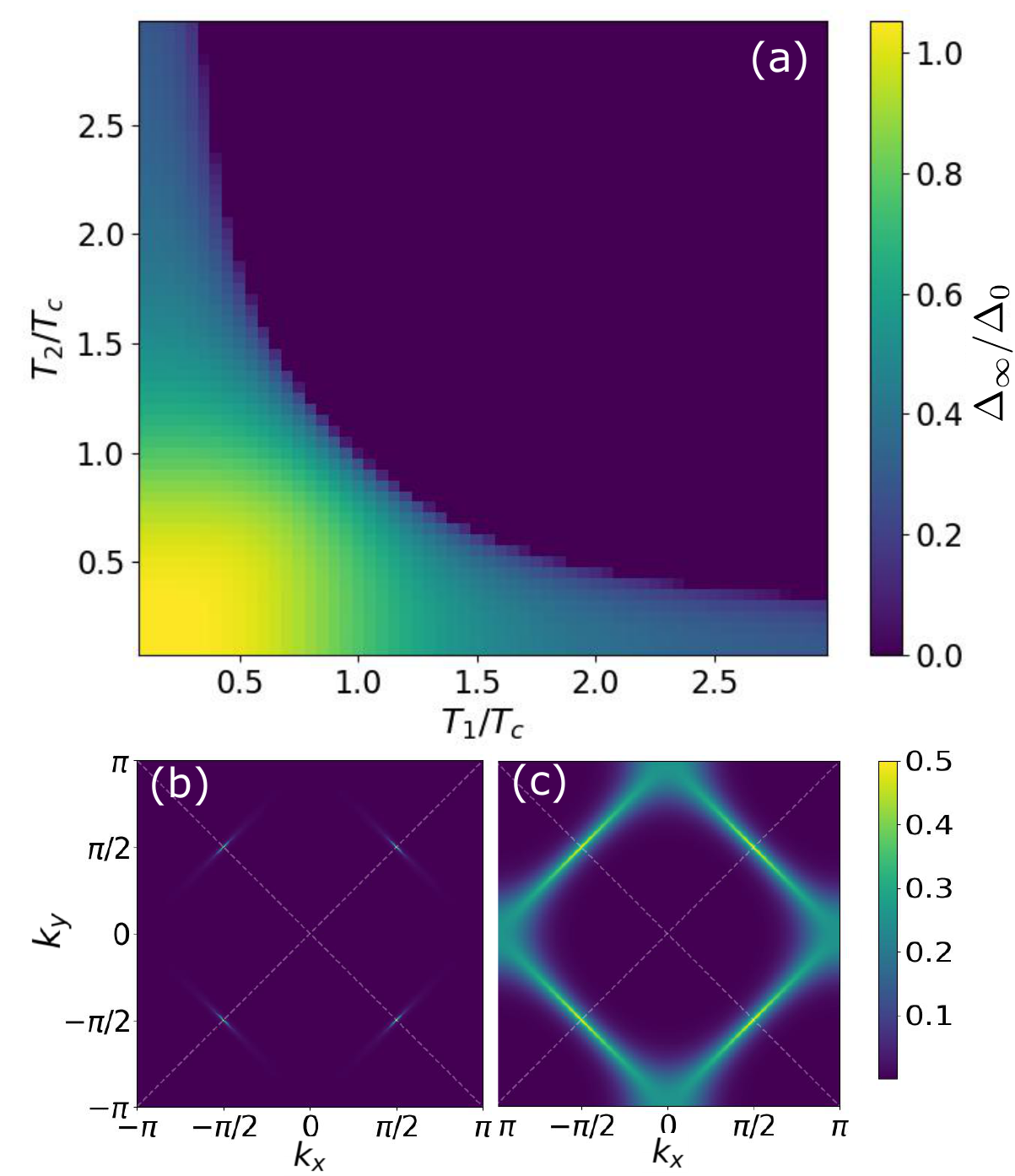}
\caption{(a) Steady-state gap $\Delta_\infty/\Delta_0$ as a function two temperature baths $T_1$ and $T_2$ for d-wave superconductor. Distribution of quasiparticles in the equilibrium at $T_1/T_c=T_2/T_c=0.1$ (b), and in the steady state with $T_1/T_c=1.7$ and $T_2/T_c=0.3$ (c).}\label{figure:fig3}
\end{figure}

{\it Results for d-wave non-equilibrium superconductivity.}
Another natural playground for the $\mathcal{O}(1)$ manipulation of superconductivity are also unconventional superconductors, e.g., with d-wave order relevant for cuprates. Since their thermodynamic properties are governed by a node in the gap, such systems could be efficiently controlled through nonthermal population distributions.

We show results for the d-wave symmetry of interaction with $\lambda(p,p')=\lambda (\cos(p_x)-\cos(p_y)) (\cos(p'_x)-\cos(p'_y))$ in Eq.~\eqref{eq:H0} and $\Delta_p = \Delta (\cos(p_x)-\cos(p_y))$. Similarly to the s-wave case, we again find large regions of temperatures with  $T_2 < T_c < T_1$ or  $T_1 < T_c < T_2$ that show a finite $\lim_{t\to\infty}\Delta(t)=\Delta_\infty$ in the steady state (Fig.~\ref{figure:fig3}a). The main difference is that the finite superconducting order exists in a much broader range of temperatures than the s-wave~(note different axis ranges). A comparison between the low-temperature equilibrium state~(Fig.~\ref{figure:fig3}b) and the highly nonthermal steady state at $T_1/T_c=1.7$ and $T_2/T_c=0.3$~(Fig.~\ref{figure:fig3}c) shows that the latter retains a finite order parameter, despite a broad quasiparticle distribution with peaks in the nodal region. The quasiparticle energy distribution (not shown) confirms a mechanism similar to that in Fig.~\ref{figure:fig2}, where the steady-state quasiparticles are shifted to higher energies compared to the thermal distribution.

{\it Conclusions.} We have demonstrated that coupling a superconductor to two distinct thermal baths can stabilize superconducting order, even when one bath is held at a temperature well above the equilibrium $T_c$. 
The core mechanism relies on a ``greenhouse effect'' where weak driving can lead to strong response in system with approximately conserved quantities \cite{lange17}. In our setup with approximately conserved quasiparticles, weak overall non-thermal dissipation induced an $\mathcal{O}(1)$ redistribution of quasiparticle occupation, thereby stabilizing order in regimes where it is absent in equilibrium.
The purpose of this work is to establish a generic framework for dissipative engineering based on the ``greenhouse effect", with the aim of manipulating symmetry-broken states using superconductivity as a paradigmatic example. While GGE approximation used in this work has been extensively tested for integrable and non-interacting Hamiltonians~\cite{lange17,lange18,reiter21,schmitt22,ulcakar23,ulcakar24,ulcakar2025,bouchoule20,rossini21,mazza22,gerbino23,perfetto23,rowlands23,riggio24,starchl22,starchl24, lumia24,lehr25,marche26}, more advanced diagrammatic tools could be used to include fluctuations beyond mean-field~\cite{matthies2018,künzel2026nessi20nonequilibriumsystems,li2020,li2021}. 

Extending this formalism to specific experimental platforms requires incorporating the spectral properties of the environment. For instance, applications to Fabry-Pérot~\cite{pini2026,curtis2019} or split-ring resonators~\cite{gao2020} will necessitate tailored modifications of the system-bath coupling. We considered a coupling to thermal baths that induces transitions across a large fraction of the Brillouin zone. Such a regime can be realized by placing the superconductor in close proximity to confined electromagnetic modes, such as surface plasmon polaritons~\cite{ashida2020,Economou1969} or polariton in hyperbolic materials~\cite{caldwell2014sub,Sternbach2021}. Alternatively, quantum simulators like cavity quantum electrodynamics architectures~\cite{Young2024,Young25,chu2025} offer a promising route to engineer the requisite bath couplings \cite{mivehvar2021}. Looking forward, an important open question is whether similar dissipative protocols can be generalized to stabilize other symmetry-broken phases, including charge-density waves~\cite{Jarc2023}, magnetic orders~\cite{kiffner2019,sentef2020,curtis2022}, or excitonic condensates~\cite{sun2024,mazza2019,davari2025,lenk2020,lenk2022,dmytruk2021gauge}.

\begin{acknowledgments}
We thank M. Eckstein, D. Fausti, M. Feigelman, V. Kabanov F. Piazza and M. Pini for useful discussions, and I. Ul\v{c}akar for collaboration in early stages of the project. Research of A. L. was funded by Google.org. Y. H. acknowledges Public Scholarship, Development, Disability and Maintenance Fund of the Republic of Slovenia, scholarship No. 11011-8/2025. 
Z.L acknowledges the support by P1-0044 program of the Slovenian Research and Innovation Agency (ARIS), European Union Horizon 2020 under the QuantERA II project QuSiED (No. 101017733), and ERC StG 2022 project DrumS by Horizon Europe, Grant Agreement 101077265. D.G. is supported by the Slovenian Research and Innovation Agency (ARIS) under Programs No. P1-0044, No. J1-2455, No. N1-0493, and No. MN0016-106, European Union Marie Sklodowska-Curie Doctoral Network SPARKLE grant No. 101169225 and ERC CoG 2026 project META-QMS by Horizon Europe, Grant Agreement 101230115.
\end{acknowledgments}

\bibliography{biblio}

\section{End Matter}

\subsection{Implementation of thermal baths}
\begin{figure*}[t!]    
\includegraphics[]{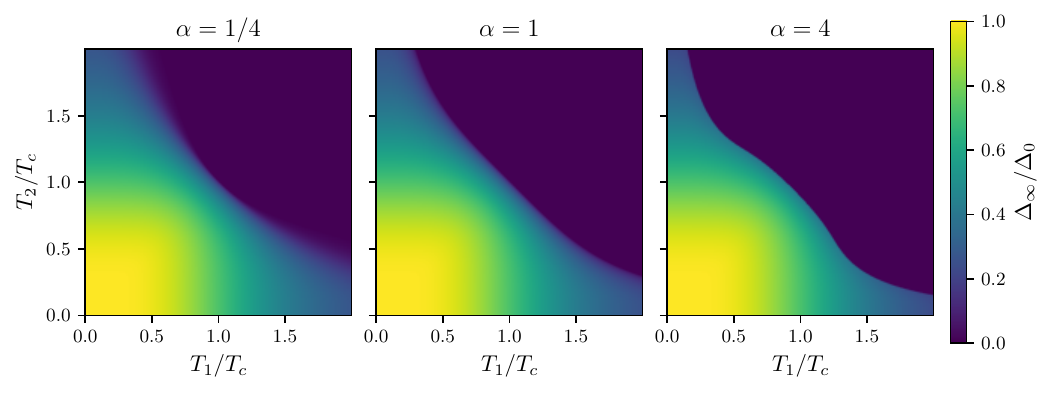}
\caption{Dependence of the steady state gap on the relative strength $\alpha=\alpha_{+-}/(\alpha_{++}+\alpha_{--})$ of scattering and two-particle creation/annihilation term for the s-wave pairing case. Results for $\alpha=1/4, 4$ are comparable to the $\alpha=1$ considered in the main text, emphasising that our results are not fine tuned to a specific choice of $\alpha$.}\label{figure:fig5}
\end{figure*}
To obey detailed balance at temperature $T$, rates of different processes in the dissipator \eqref{eq:thermalD} are bound to satisfy certain conditions \cite{Zanoci23}. Namely, two particle creation and annihilation are related by
$e^{-( \epsilon_{k_1} +  \epsilon_{k_2})/T} \gamma^{(-,-)}_{k_1,k_2} = \gamma^{(+,+)}_{k_2,k_1}$.
Scattering must satisfy
$e^{ ( \epsilon_{k_1} - \epsilon_{k_2})/T} \gamma_{k_1,k_2}^{(+,-)} = \gamma_{k_2,k_1}^{(+,-)}$, but is in principle independent of two particle annihilation/creation. 
A priori, $\gamma^{(+,+)}_{k_1,k_2} = \gamma^{(+,+)}_{k_2,k_1}$ and $\gamma^{(-,-)}_{k_1,k_2} = \gamma^{(-,-)}_{k_2,k_1}$. We resort to the choice
\begin{equation}
\label{eq:gammas}
\gamma^{(s_1,s_2)}_{k_2,k_1}(T) = \alpha_{s_1,s_2} \, f_{s_1}\left(\frac{\epsilon_{k_1}}{T}\right) f_{s_2}\left(\frac{\epsilon_{k_2}}{T}\right)
\end{equation}
with $f_s\left(\frac{\epsilon_{k}}{T}\right) = \frac{1}{1+e^{ s \epsilon_k/T}}$ the Fermi occupation at temperature $T$ for $s=+$. 
We also introduce a parameter that quantifies the imbalance between the two types of dissipators:
$\alpha =\alpha_{+-}/(\alpha_{++} + \alpha_{--})$.
Note that the above constraints imply $\alpha_{++} = \alpha_{--}$.

In our modelling of the baths, the relative strength $\alpha$ of the scattering and two-particle creation/annihilation terms is not fixed, since any choice of $\alpha$ yields a thermal steady state. In Fig.~\ref{figure:fig5} we compare the phase diagrams at different $\alpha=1/4, 1, 4$, confirming that the main results of our study are not strongly dependent on a specific choice of $\alpha$. A finite $\alpha\sim\mathcal{O}(1)$ is anticipated in general since a generic scattering/annihilation/creation of electrons would result in the presence of all possible processes in terms of Bogoliubov operators. In the main text, all results correspond to $\alpha=1$.

\onecolumngrid

\subsection{Steady state occupation number}
\label{sec: Steady state occupation number.}
Here we show how the steady-state condition can be simplified using the structure of Eq.~(\ref{eq:gammas}). We first introduce
\begin{align}
\label{eq:AB}
     A(T_j)\equiv  \sum_{k^\prime}  \left[ f_+ \left(\frac{\epsilon_{k^\prime}}{T_j}\right) (1 - n_{k^\prime})+ \alpha\,f_-\left(\frac{\epsilon_{k^\prime}}{T_j}\right)n_{k^\prime} \right], \nonumber\\ B(T_j) = \sum_{k^\prime}\left[  f_-\left(\frac{\epsilon_{k^\prime}}{T_j}\right)  n_{k^\prime}  + \alpha\,f_+\left(\frac{\epsilon_{k^\prime}}{T_j}\right)(1 - n_{k^\prime}) \right].
\end{align}
Using this notation, the steady-state condition $\partial_t n_k = 0$ following from Eq.~(\ref{eq:scattering}) can be written as
\begin{equation}
    0  = 2\alpha_{++}\gamma_0\sum_{j = 1,2} \left\{f_+\left(\frac{\epsilon_k}{T_j}\right)(1 - n_{k}) A(T_j)    - f_-\left(\frac{\epsilon_k}{T_j}\right)n_{k}B(T_j)\right\}.
\end{equation}
Solving this equation for $n_k$, we obtain
\begin{equation}
\label{eq:steady_state_ocupation}
    n_{k} = \frac{f_+\left(\frac{\epsilon_k}{T_1}\right) A(T_1) + f_+\left(\frac{\epsilon_k}{T_2}\right) A(T_2)}{f_+\left(\frac{\epsilon_k}{T_1}\right) A(T_1) + f_+\left(\frac{\epsilon_k}{T_2}\right) A(T_2) + f_-\left(\frac{\epsilon_k}{T_1}\right) B(T_1) + f_-\left(\frac{\epsilon_k}{T_2}\right) B(T_2)}.
\end{equation}
Relation~(\ref{eq:steady_state_ocupation}), together with Eq.~(\ref{eq:AB}) and the self-consistency equation for the gap, Eq.~(\ref{eq:Delta_t}), form a closed system of equations that determine the steady state. This system can be solved numerically using an iterative scheme. 

Alternatively, self-consistent equations for the d-wave superconductor were solved using the time propagation technique described in the main text.

\subsection{Phase transition line for  $\alpha = 1$}
The system can be further simplified in the special case $\alpha = 1$ ($\alpha_{++} = 1$ and $\alpha_{+-} = 2$). For this choice, one finds that $A(T) = B(T)$, which leads to
\begin{equation}
    n_k = \frac{f_+\left(\frac{\epsilon_k}{T_1}\right) + F  f_+\left(\frac{\epsilon_k}{T_2}\right)}{1 +F}, \quad F \equiv \frac{A(T_2)}{A(T_1)},
\end{equation}
In this special case, the occupation number is simply a linear combination of distributions corresponding to two different temperatures. This allows us to derive a self-consistent equation for $F$:

\begin{equation}
    F = \frac{N_{+,-}^{2,1} + F  N_{+,-}^{2,2} + N_{-,+}^{2,1} + F  N_{-,+}^{2,2}}{N_{+,-}^{1,1} + F  N_{+,-}^{1,2} + N_{-,+}^{1,1} + F  N_{-,+}^{1,2}}.
\end{equation}
Here we have introduced the notation
\begin{equation}
    N_{s_1,s_2}^{ij} \equiv \sum_{k} f_{s_1}\left(\frac{\epsilon_k}{T_i}\right)f_{s_2}\left(\frac{\epsilon_k}{T_j}\right).
\end{equation}
The equation for $F$ can be solved explicitly. As a result, the problem reduces to a single self-consistency equation for the gap. This equation can be further simplified in the vicinity of the transition line, where $\Delta \to 0$. In this limit, the self-consistency equation takes the form
\begin{equation}
         \frac{1}{1+F}\log\frac{T_c}{T_1 } + \frac{F}{1 +F }\log\frac{T_c}{T_2 }  =  \frac{7 \zeta(3)}{8 \pi^2}\left(\frac{1}{1+F}\left(\frac{\Delta}{T_1}\right)^2  + \frac{F}{1+F}\left(\frac{\Delta}{T_2}\right)^2 \right).
\end{equation}

\begin{figure}[t!]    
\includegraphics[width=0.4\columnwidth]{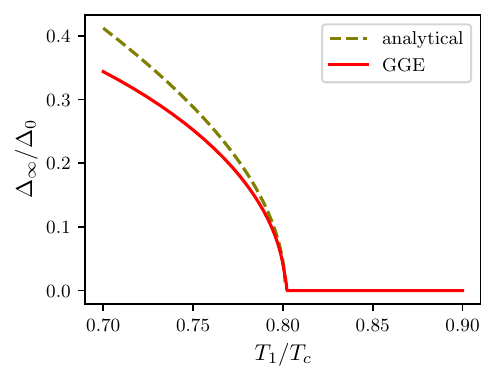}
\caption{The steady state gap in a system interacting with two baths as a function of the temperature of the first bath, while the temperature of the second bath is fixed at $T_2/T_c = 1.2$. The solid line represents the prediction of the GGE. The dashed line shows the asymptotic analytical solution in the regime $\Delta_0 \gg \Delta_{\infty}$.}\label{figure:fig6}
\end{figure}

In evaluating $F$, one should use the gapless spectrum. The main consequence of the above equation is the emergence of a square-root singularity, which reflects the second-order nature of the phase transition in our dissipative setup. The line in the  $(T_1,T_2)$ plane along which the left-hand side of the above equation vanishes defines the phase transition boundary.
Fig.~\ref{figure:fig6} shows comparison of the numerical (GGE) and the analytical expansion expression.

\clearpage

\renewcommand{\thetable}{S\arabic{table}}
\renewcommand{\thefigure}{S\arabic{figure}}
\renewcommand{\theequation}{S\arabic{equation}}
\renewcommand{\thepage}{S\arabic{page}}
\renewcommand{\thesection}{S\arabic{section}}

\onecolumngrid

\setcounter{figure}{0}
\setcounter{equation}{0}
\setcounter{page}{0}

\begin{center}
{\large \bf Supplemental Material:\\
Non-equilibrium dissipative stabilization of  s- and d-wave superconductivity}\\
\vspace{0.3cm}
Aleksey Lundkin$^{1}$, Yury Holubeu$^{1}$, Denis Gole\v{z}$^{2,3}$, and Zala Lenar\v ci\v c$^{2}$\\
$^1${\it Nanocenter, 1000 Ljubljana, Slovenia} \\
$^2${\it Jo\v{z}ef Stefan Institute, 1000 Ljubljana, Slovenia} \\
$^3${\it Faculty for Mathematics and Physics, University of Ljubljana, Jadranska ulica 19, 1000 Ljubljana, Slovenia} \\
\end{center}




\label{pagesupp}

\section{General intrinsic scattering rate}

\begin{figure*}[b!]
\centering
\makebox[\textwidth][c]{%
\includegraphics[width=0.48\textwidth]{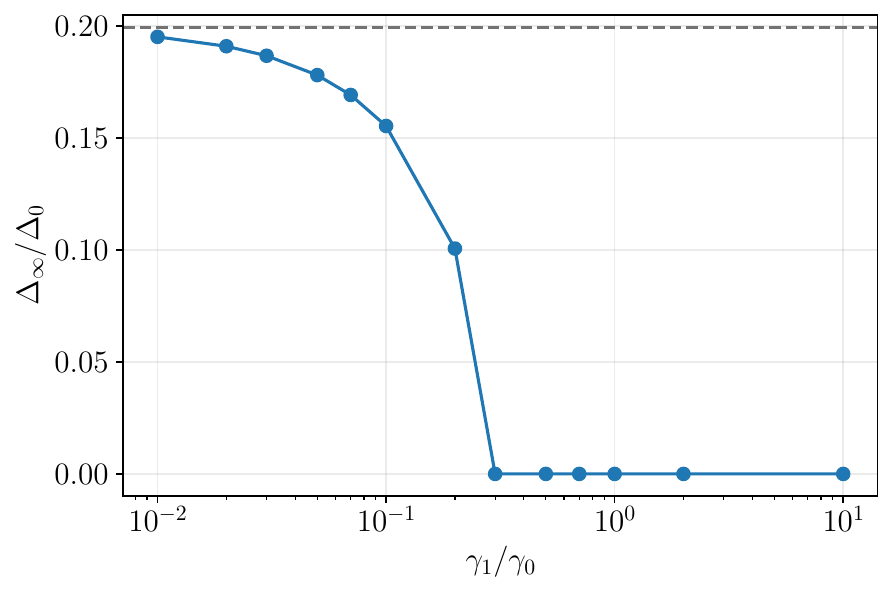}%
\includegraphics[width=0.48\textwidth]{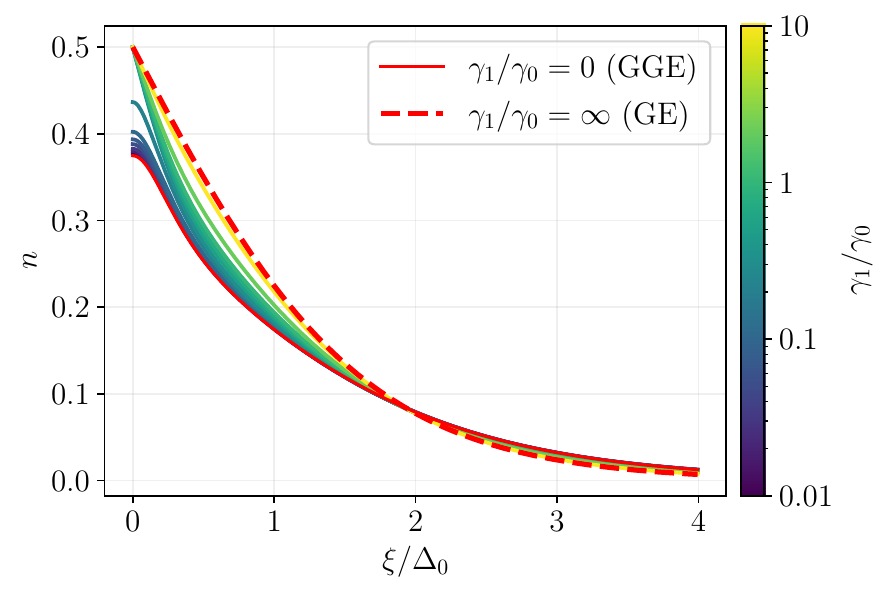}%
}
\caption{(Left) Dependence of the steady-state gap $\Delta_\infty$ on the ratio $\gamma_1/\gamma_0$. Dashed line represent the $\gamma_1=0$ result. (Right)~Occupation numbers as a function of energy for various values of $\gamma_1/\gamma_0$. Results interpolate between the GGE and the thermal GE functions. Parameters: $T_1/T_c = 0.3$ and $T_2/T_c = 1.8$.}
\label{fig:general_scattering}
\end{figure*}

In the main text, we considered only the two extreme cases of Eq.~\eqref{eq:scattering} with negligible or strong intrinsic scattering $\gamma_1$.  
In Fig.~\ref{fig:general_scattering}, we demonstrate the influence of a finite scattering rate $\gamma_1$, phenomenologically pushing the system towards a thermal distribution at a  temperature determined via the stationary energy exchange with the baths, Eq.~\eqref{eq:Hrate}. As anticipated in the main text, in the limit $\gamma_1 \ll \gamma_0$, the occupation numbers are well described by the generalized Gibbs ensemble (GGE) solution. In the opposite limit, $\gamma_1 \gg \gamma_0$, the system approaches the thermal Gibbs ensemble (GE) behaviour with zero gap at deliberately chosen $T_1/T_c=0.3$ and $T_2/T_c=1.8$.

\section{Details of the numerical simulation}

For a given set of bath temperatures at s-wave pairing, we used the following iterative algorithm to determine the steady-state quasiparticle distribution and the superconducting gap.
We start from an initial guess for the quasiparticle occupation and the gap. First, we compute the parameters $A(T_{i})$ and $B(T_{i})$, $i=1,2$, according to Eq.~(\ref{eq:AB}). Next, we update the quasiparticle distribution according to Eq.~(\ref{eq:steady_state_ocupation}). Using the updated distribution, we then calculate a new value of the gap from the self-consistency equation
\begin{equation}
    \ln\left(\frac{\Delta_0}{\Delta_{\mathrm{new}}} \right)
    =
    2\int_0^{\infty}
    \frac{n(\epsilon(\xi))\, d\xi}{\epsilon(\xi)},
    \qquad
    \epsilon(\xi) = \sqrt{\xi^2 + \Delta_{\mathrm{old}}^2}.
\end{equation}
For the numerical integration, we used the cutoff
$\Lambda = 10 \Delta_0$ and discretized the interval with
$L = 1000$ points. The iterative procedure was terminated when the change in the gap between successive iterations became smaller than
$10^{-4}\Delta_0$.

The calculation of steady state distribution for the Hamiltonian with d-wave pairing has been obtained from a time propagation in which we initialize the quasiparticle distribution in some state, update the quasiparticle occupation numbers $n_{p,\sigma}(t)$ following Eq.~\eqref{eq:scattering} and adjust the gap $\Delta_p(t)$ at each time step according to Eq.~\eqref{eq:Delta_t} until it converges. We discretize the two dimensional  Brillouin zone with $L\times L=1600 \times 1600$ momentum points and set the cutoff $\omega_D=0.6$ in the units of fermion hopping $t_0$, $\xi_p=-2 t_0 (\cos(k_x)+\cos(k_y))$.

\end{document}